\documentclass[11pt]{article}
\usepackage[pdftex]{graphicx,color} 
\usepackage{jheppub}
\usepackage{amsmath}
\usepackage{amssymb}
\usepackage{comment}
\usepackage{multirow}
\usepackage{mathtools}
\usepackage{dsdshorthand}
\usepackage{mleftright}
\newcommand{\bea}{\begin{equation}\begin{aligned}}
\newcommand{\eea}[1]{\label{#1}\end{aligned}\end{equation}}
\newcommand{\beq}{\begin{equation}}
\newcommand{\eeq}{\end{equation}}
\newcommand   \xb  {\bar{x}}
\newcommand   \zb  {\bar{z}}

\newcommand\al{{\alpha}}
\newcommand   \ab  {{\bar{\alpha}}}

\newcommand{\cN}{{\mathcal N}}

\usepackage{tikz}
\usetikzlibrary{arrows,calc,shapes,decorations.pathmorphing,decorations.markings}
\tikzset{
>=stealth',
help lines/.style={dashed, thick},
axis/.style={<->},
important line/.style={thick},
connection/.style={thick, dotted},
->-/.style={decoration={
  markings,
  mark=at position #1 with {\arrow{>}}},postaction={decorate}},
  twopt/.style={
    circle,
    draw,
    fill=black,
    inner sep=1pt,
    minimum size=1pt
  },
  spinning/.style={
    thick,
    postaction={
      decorate,
      decoration={
        markings,
        mark=at position 0.5 with {\arrow{>}}
      }
    }
  },
}
\newcommand{\diagramEnvelope}[1]{#1}

\title{On the spectrum and structure constants of short operators in N=4 SYM at strong coupling}

\author{Luis F. Alday, Tobias Hansen and Joao A. Silva}

\affiliation{Mathematical Institute, University of Oxford,
Woodstock Road, Oxford, OX2 6GG, UK}

\abstract{
We study short operators in planar $\mathcal{N}=4$ SYM at strong coupling, for general spin and $SO(6)$ symmetric traceless representations. At strong coupling their dimension grows like $\Delta \sim 2\sqrt{\delta} \lambda^{1/4}$ and their spectrum of degeneracies can be analysed by considering the massive spectrum of type II strings in flat space-time.  We  furthermore compute their structure constants with two arbitrary chiral primary operators. This is done by considering the four-point correlator of arbitrary chiral primary operators at strong coupling in planar $\mathcal{N}=4$ SYM, including the supergravity approximation plus the infinite tower of stringy corrections that contributes in the flat space limit. Our results are valid for generic rank $n$ symmetric traceless representations of $SO(6)$ and in particular for $n \gg 1$, as long as $n \ll \lambda^{1/4}$.
}

\emailAdd{alday@maths.ox.ac.uk, tobias.hansen@maths.ox.ac.uk, joao.silva@maths.ox.ac.uk}

\begin{document}
\maketitle

 \section{Introduction}

It has been known since the work of Gubser, Klebanov and Polyakov that massive string states in $AdS_5 \times S^5$ couple to short single-trace operators in $\mathcal{N}=4$ SYM  \cite{Gubser:1998bc}. These operators have conformal dimensions which grow as $\Delta \sim 2\sqrt{\delta} \lambda^{1/4}$ at strong t'Hooft coupling $\lambda$, where $\delta=1,2,\ldots$ is the string mass level.
In this paper we revisit these operators $\cO_{\delta,\ell,{\cal R}}$ for general spin $\ell$ and in the $SO(6)$ representation ${\cal R}$.

One of the reasons we are interested in these operators is their appearance in stringy correlators.
In \cite{Alday:2022uxp,Alday:2022xwz} we recently derived the full $1/\sqrt{\lambda}$ correction to the Virasoro-Shapiro amplitude in $AdS_5 \times S^5$, i.e.\ the correlator $\< \cO_2 \cO_2 \cO_2 \cO_2 \>$ at large $N$ and strong coupling including all stringy corrections, where $\cO_2$ is the chiral primary operator in the stress-tensor multiplet.
From this result one can extract averaged corrections to conformal dimensions and structure constants for all the massive short operators $\cO_{\delta,\ell,[0,0,0]}$.
A natural question is whether these averages can be unmixed by considering more general correlators $\< \cO_{p_1} \cO_{p_2} \cO_{p_3} \cO_{p_4} \>$, where $\cO_{p_i}$ is a chiral primary of dimension $p_i$.\footnote{A similar unmixing was carried out for light double trace operators in supergravity in \cite{Alday:2017xua,Aprile:2017bgs,Aprile:2017xsp}.} As we will show in this paper, unfortunately this is not the case. 

Another motivation to study short operators is to further bridge the gap between the conformal bootstrap and
integrability. Explicit integrability results for dimensions of short operators at strong coupling are currently available for those operators who sit on the leading Regge trajectory \cite{Gromov:2011de,Basso:2011rs,Gromov:2011bz}, including the familiar Konishi operator. These results agree with \cite{Alday:2022uxp,Alday:2022xwz}. One  advantage of our approach however, is that we can compute structure constants, for which integrability techniques are still under development, see however \cite{Basso:2022nny} for recent progress relevant for the case at hand. 

To understand the spectrum of short massive operators at strong coupling,
we relate it
to the spectrum of type II strings in flat space, compactified on $S^5$.
The relation between conformal dimensions and masses is the one from \cite{Gubser:1998bc}
and the representation theory for spin and R-symmetry is similar to the weak coupling treatment of \cite{Bianchi:2003wx,Beisert:2003te}.
One result of this analysis is that operators on the leading Regge trajectory $\ell = 2 \delta-2$ are non-degenerate while operators on the next Regge trajectory $\ell = 2 \delta-3$ have degeneracy 2.

By combining the precise result for the supergravity correlator $\< \cO_{p_1} \cO_{p_2} \cO_{p_3} \cO_{p_4} \>$ from \cite{Rastelli:2016nze,Aprile:2018efk} with the flat space limit formula of \cite{Penedones:2010ue,Fitzpatrick:2011hu} we derive the leading structure constants for the coupling of two arbitrary chiral primaries with the massive short operator $\cO_{\delta,\ell,[n,0,0]}$. This result is valid for generic R-charges of the three operators provided $p_i \ll \lambda^\frac14$.
This restriction implies that in the flat space limit, all momenta in the $S^5$ directions vanish and all $SO(6)$ representations are KK-modes of the $\mathbb{R}^5$ singlet. One could also study the case $p_i \sim \lambda^\frac14$ by considering a different flat space limit that acts non-trivially on $S^5$, see \cite{Aprile:2020luw}.

The structure constants of two chiral primaries and one short massive operator on the leading Regge trajectory have previously been computed in \cite{Minahan:2012fh,Bargheer:2013faa,Minahan:2014usa} and our computation reproduces this result in this special case.
One simplification of our computation compared to \cite{Minahan:2012fh,Bargheer:2013faa,Minahan:2014usa} is that at each step we work at the level of long supersymmetry multiplets. Since the long superconformal multiplet must become the usual long multiplet in the flat space limit, we strip off the corresponding factor at each step: when analysing the spectrum and when considering partial wave decompositions in flat space and AdS.

This paper is organised as follows.
In section \ref{sec:spectrum} we study the massive spectrum of type II superstrings compactified on $S^5$, which we expect to coincide with the spectrum of the corresponding massive operators in $\mathcal{N}=4$ SYM at strong coupling.
In section \ref{sec:partial_wave_flat} we review the corresponding partial wave decomposition of the Virasoro-Shapiro amplitude in flat space and
in section \ref{sec:partial_wave_ads} we use the flat space limit of the conformal partial wave decomposition to compute the structure constants of two chiral primaries and one short massive operator.
Section \ref{sec:mixing} discusses the question whether the general structure constants allow us to unmix OPE data from the AdS Virasoro-Shapiro amplitude.
In section \ref{sec:comparison} we compare our structure constants to the previously known result for the special case of operators on the leading Regge trajectory.
We conclude in section \ref{sec:conclusion}.

\section{Spectrum of massive strings}
\label{sec:spectrum}

In this section we study the spectrum of short operators at strong coupling. The simplest example is the Konishi operator whose conformal dimension at large $\lambda$ is given by
\beq
\Delta_{\cal K} = 2 \lambda^{1/4} -2 +\frac{2}{\lambda^{1/4}}+  \cdots.
\eeq
More generally, short operators at strong coupling are labelled by their 'level' $\delta$, spin $\ell$ and $SO(6)$ representation ${\cal R}$. At strong coupling and for small quantum numbers $\ell,{\cal R}$ (hence the name short\footnote{Despite being called short, these operators are in long superconformal multiplets.}) compared to $\lambda^{1/4}$, their dimension grows as 
\beq
\Delta_{\delta,\ell,{\cal R}} = 2 \sqrt{\delta} \lambda^{1/4} +  \cdots.
\label{dimension_levels}
\eeq
The question we would like to answer is that of their degeneracy for a given spin $\ell$ and $SO(6)$ representation. Following \cite{Gubser:1998bc} this maps to a computation of the massive spectrum of type II string theory in flat space. More precisely, the massive spectrum of type II strings on $\mathbb{R}^{1,4}\times S^5$ comes in representations of $SO(4)\times SO(6)$ and is expected to match the spectrum of short operators at strong coupling. The masses in flat space $m$ are directly related to the conformal dimensions \eqref{dimension_levels}
under identification of the quadratic Casimirs of $SO(4,2)\times SO(6)$ and the 10d Poincare group in the limit where the AdS radius squared $R^2 = \alpha' \sqrt{\lambda}$ is very large
\beq
-\Delta^2 = -R^2 m^2 + \ldots\,.
\label{casimir_massive}
\eeq
The massive spectrum of type II string theory in flat space is given by 
\beq
m^2 = \frac{4 \delta}{\alpha'}\,, \qquad \delta = 1,2,3,\ldots\,.
\label{mass_levels}
\eeq

At each string level $\delta$ the $SO(9)$ (the massive little group for $\mathbb{R}^{1,9}$) representations are encoded in the character polynomial
\beq
T_\delta  = T_1 \otimes ( \text{vac}_\delta \otimes \text{vac}_\delta )\,,
\label{T_flat}
\eeq
where $T_1$ is the long multiplet in flat space\footnote{We label irreducible representations of  $SO(d)$ (and their characters) by $[a_1,\ldots,a_r]$, where $r$ is the rank of $SO(d)$ and $a_i$ are non-negative integers which give the coefficients in the decomposition of the highest weight vectors into fundamental weights. Similarly irreducible representations of $SO(d_1)\times SO(d_2)$ are labelled by $[a_1,\ldots,a_{r_1};b_1,\ldots,b_{r_2}]$.}
\beq
T_1 = \left([2,0,0,0]+[0,0,1,0]+[1,0,0,1]\right)^2 =
\left({\bf 44+84+128}\right)^2 \,,
\eeq
and $\text{vac}_\delta$ is given explicitly up to $\delta=7$ in \cite{Bianchi:2003wx} and a generating function is given in \cite{Hanany:2010da}.
The first few cases are
\bea
{\rm vac}_1 ={}& [0, 0, 0, 0] = {\bf 1}\,,\\
{\rm vac}_2 ={}& [1, 0, 0, 0] = {\bf 9}\,,\\
{\rm vac}_3 ={}&      [2, 0, 0, 0]+[0, 0, 0, 1] =    {\bf 44}+{\bf 16}\,,\\
{\rm vac}_4 ={}&
[3, 0, 0, 0]+[1, 0, 0, 1]+[0, 1, 0, 0]+[1, 0, 0, 0]
    = {\bf 156} +{\bf 128} + {\bf 36}+{\bf 9}\,.
\eea{vac}
In order to extract from this the spectrum of massive string operators in $\text{AdS}_5 \times \text{S}^5$ we will mostly follow \cite{Bianchi:2003wx,Beisert:2003te}.\footnote{A similar analysis was done for $SO(6)$ singlets in \cite{Antunes:2020pof}.}

The first step is to replace the flat long multiplet $T_1$ in \eqref{T_flat} by the long superconformal multiplet $\mathcal{T}_\text{sconf}$. However, as we really want to study the spectrum of superconformal primaries, we will not multiply by $\mathcal{T}_\text{sconf}$. By considering only ${\rm vac}_\delta^2$ we are directly counting superconformal primaries. We have\footnote{We use SageMath \cite{sagemath} to compute tensor products and branchings.}
\bea
{\rm vac}_1^2 ={}& [0, 0, 0, 0]\,,\\
{\rm vac}_2^2 ={}& [2, 0, 0, 0] + [0, 1, 0, 0] + [0, 0, 0, 0]\,,\\
{\rm vac}_3^2 ={}& [4,0,0,0]+[2,1,0,0]+ 2[2,0,0,1]+[0,2,0,0]+[2,0,0,0]+ 2[1,0,0,1]\\
&+[0,0,0,2]+[0,0,1,0]+ 2[0,1,0,0]+[1,0,0,0]+ 2[0,0,0,0]\,.
\eea{vac^2}
Next we branch the $SO(9)$ representations into irreducible representations of $SO(4)\times SO(5)$, corresponding to the split into $AdS_5$ and $S^5$ directions. For the first mass levels this gives
\begin{align}
{\rm vac}_1^2 ={}& [0, 0; 0, 0]\,,\label{branching}\\
{\rm vac}_2^2 ={}&[2,2;0,0]
+ [2,0;0,0]
+ [0,2;0,0]
+2[0,0;0,0]
+2[1,1;1,0]
+ [0,0;2,0]
+ [0,0;0,2]\,.
\nonumber
\end{align}
The full expressions get very lengthy very quickly, but simplify if one looks at specific representations.\footnote{For instance, we can focus on symmetric traceless representations.
These are the $[\ell,\ell]$ of $SO(4)$ and the $[m,0]$ of $SO(5)$.
Projecting onto these representations we get
\bea
{\rm vac}_1^2 |_{STT} ={}& [0, 0; 0, 0]\,,\\
{\rm vac}_2^2 |_{STT} ={}& [2,2;0,0]
+2[0,0;0,0]
+2[1,1;1,0]
+ [0,0;2,0]\,,\\
{\rm vac}_3^2 |_{STT} ={}&
 [4,4;0,0]+
 4[2,2;0,0]+
 2[1,1;0,0]+
6[0,0;0,0]+
 2[3,3;1,0]+
 8[1,1;1,0]\\
&+ 2[0,0;1,0]+
 3[2,2;2,0]+
 4[0,0;2,0]+
 2[1,1;3,0]+
 [0,0;4,0]\,.
\eea{branching_STT}}

To get the final degeneracy in terms of $SO(6)$ representations, we compactify five dimensions into $\text{S}^5$ \cite{Salam:1981xd}.
This replaces each $SO(5)$ irrep $[m,n]$ by a Kaluza-Klein tower of $SO(6)$ representations \cite{Bianchi:2003wx}
\beq
{\rm KK}_{[m,n]} =
\sum_{r=0}^{m} 
\sum_{s=0}^{n} 
\sum_{p=m-r}^\infty 
\left[p,\,r+ n- s,\, r+ s\right]
+\sum_{r=0}^{m-1} 
\sum_{s=0}^{n-1} 
\sum_{p=m-r-1}^\infty
\left[p,\,r+ n- s ,\, r+ s+1 \right]
\,.
\label{KK}
\eeq

\subsection{Massive strings coupling to two chiral primaries}
\label{sec:spectrum_correlator}

In the remainder of the paper we will study the structure constants of one short massive operator and two chiral primary operators to leading order in $1/\lambda$, so it is important to understand which part of the massive spectrum admits such couplings.
To this end we need to refine the matching of Casimirs which we did for short massive operators in \eqref{casimir_massive}.
A general operator with dimension $\Delta$, spin $\ell$ and $SO(6)$ representation $[n,0,0]$ has the $SO(4,2)$ Casimir
\beq
c^2_{SO(4,2),\Delta,\ell} = \Delta(4-\Delta) + \ell (\ell+4-2)\,,
\eeq
and the $SO(6)$ Casimir
\beq
c^2_{SO(6),[n,0,0]} = n(n+6-2)\,.
\eeq
In the flat space limit the $SO(6)$ Casimir should equal the Casimir for the five spacelike dimensions that are compactified into $S^5$, i.e.\ 
we can split the 10d momentum $k^M$ in flat spacetime with metric $(-+\ldots+)$ into
\beq
k^M = (k^\mu,k^m)\,, \qquad \mu=0,\ldots 4\,, \quad m =5\ldots 9\,,
\label{ksplit}
\eeq
and assign separately for the limit $R\to \infty$ (where $R$ is the radius of both $AdS_5$ and $S^5$)
\bea
c^2_{SO(4,2),\Delta,\ell} &= R^2 k_\mu k^\mu + \ldots\,,\\
c^2_{SO(6),[n,0,0]}  &= R^2 k_m k^m + \ldots\,.
\eea{casimirs_split}
For massive operators this agrees with the assignment \cite{Minahan:2012fh,Bargheer:2013faa}.

Chiral primary operators $\cO_p$ are the superconformal primaries of $1/2$-BPS superconformal multiplets and their quantum numbers are fixed in terms of a single integer $p\geq 2$, namely $\Delta=p$, $\ell=0$ and their $SO(6)$ representation is $[p,0,0]$.
\eqref{casimirs_split} implies that in the flat space limit the components $k^m$ of the momentum vanish
\beq
k^m = O\left( \frac1R \right)\,.
\label{km_small}
\eeq
This is familiar for the flat space limit of correlators of chiral operators in supersymmetric conformal field theories \cite{Chester:2018dga,Alday:2020dtb}. As we will see in section \ref{sec:partial_wave_flat}, \eqref{km_small} implies that
 to leading order in $1/\lambda$ only massive operators from the singlet of $SO(5)$ couple to two chiral primaries.
For this reason we next analyse this part of the massive spectrum in some more detail.

\subsection{Spectrum from the $SO(5)$ singlet}

We show the spectrum that arises from the singlet of $SO(5)$ separately for even and odd spins in tables \ref{SO5-00-even} and \ref{SO5-00-odd}, where we use the shorthand notation $[\ell]\equiv [\ell,\ell]$ for traceless symmetric $SO(4)$ representations of spin $\ell$.
For each entry in these tables there is a tower of KK modes in the $SO(6)$ representations
\beq
\text{KK}_{[0,0]} = \sum\limits_{q=0}^\infty [q,0,0]\,.
\label{KK00}
\eeq
We observe that the multiplicity for superconformal primaries on the leading even spin Regge trajectory is 1 and on the next even spin Regge trajectory it is 6 for all but the first few operators.
\begin{table}[h!]
\centering
\begin{tabular}{ |c|l| } 
 \hline
 $\delta$ & $\text{vac}_\delta^2$ projected to $SO(4)|_\text{even rank STT} \times SO(5)|_{[0,0]}$ \\ 
 \hline
1 & [0] \\ 
2 & [2] + 2[0] \\
3 & [4] + 4[2] + 6[0]\\
4 & [6] + 6[4] + 24[2] + 22[0] \\
5 & [8] + 6[6] + 40[4] + 157[2] + 99[0] \\
6 & [10]+ 6[8] + 52[6] + 331[4] + 1104[2] + 547[0] \\
7 & [12]+ 6[10]+ 58[8] + 461[6] + 2570[4] + 7365[2] + 3112[0] \\
 \hline
\end{tabular}
\caption{The even spin spectrum arising from the $[0,0]$ of $SO(5)$. Each entry corresponds to the KK tower of $SO(6)$ representations \eqref{KK00}.}
\label{SO5-00-even}
\end{table}
\begin{table}[h!]
\centering
\begin{tabular}{ |c|l| } 
 \hline
 $\delta$ & $\text{vac}_\delta^2$ projected to $SO(4)|_\text{odd rank STT} \times SO(5)|_{[0,0]}$ \\ 
 \hline
1 & 0 \\ 
2 & 0 \\
3 & 2[1]\\
4 & 8[3]+24[1] \\
5 & 12[5]+80[3]+172[1] \\
6 & 14[7]+128[5]+664[3]+1154[1] \\
7 & 14[9]+156[7]+1120[5]+4868[3]+7214[1] \\
 \hline
\end{tabular}
\caption{The odd spin spectrum arising from the $[0,0]$ of $SO(5)$. Each entry corresponds to the KK tower of $SO(6)$ representations \eqref{KK00}.}
\label{SO5-00-odd}
\end{table}

\subsection{Spectrum from the $SO(5)$ vector}
\label{sec:so5_vector}

It is instructive to also have a closer look at the contributions of the vector $[1,0]$ of $SO(5)$, even though they are not exchanged in the four-point functions of chiral primary operators at leading order in $1/\lambda$.
We show this part of the spectrum in tables \ref{SO5-10-even} and \ref{SO5-10-odd}, where each entry comes with the KK tower
\beq
\text{KK}_{[1,0]} = \sum\limits_{q=1}^\infty [q,0,0] + \sum\limits_{q=0}^\infty [q,1,1]\,.
\label{KK10}
\eeq
There are several important observations to make. The first one is that the only $SO(6)$ representation that is entirely described by the singlet of $SO(5)$ is the singlet $[0,0,0]$.
Further, the leading Regge trajectory with $\ell = 2 \delta - 2$ (see table \ref{SO5-00-even}) receives only contributions from the singlet of $SO(5)$.
However, the leading odd spin Regge trajectory is $\ell = 2 \delta - 3$ (see table \ref{SO5-10-odd}) and arises entirely from the vector of $SO(5)$. The degeneracy for operators on this trajectory is 2.
\begin{table}[h!]
\centering
\begin{tabular}{ |c|l| } 
 \hline
 $\delta$ & $\text{vac}_\delta^2$ projected to $SO(4)|_\text{even rank STT} \times SO(5)|_{[1,0]}$ \\ 
 \hline
1 & 0 \\ 
2 & 0 \\
3 & 2[0]\\
4 & 2[4] + 24[2] + 21[0] \\
5 & 2[6] + 46[4] + 244[2] + 157[0] \\
6 & 2[8] + 60[6] + 506[4] + 2054[2] + 1095[0] \\
7 & 2[10]+ 62[8] + 692[6] + 4689[4] + 15505[2] + 7055[0] \\
 \hline
\end{tabular}
\caption{The even spin spectrum arising from the $[1,0]$ of $SO(5)$. Each entry corresponds to the KK tower of $SO(6)$ representations \eqref{KK10}.}
\label{SO5-10-even}
\end{table}
\begin{table}[h!]
\centering
\begin{tabular}{ |c|l| } 
 \hline
 $\delta$ & $\text{vac}_\delta^2$ projected to $SO(4)|_\text{odd rank STT} \times SO(5)|_{[1,0]}$ \\ 
 \hline
1 & 0 \\ 
2 & 2[1] \\
3 & 2[3]+8[1] \\
4 & 2[5]+14[3]+46[1] \\
5 & 2[7]+16[5]+126[3]+330[1] \\
6 & 2[9]+18[7]+188[5]+1172[3]+2382[1] \\
7 & 2[11]+18[9]+228[7]+1938[5]+9714[3]+16210[1] \\
 \hline
\end{tabular}
\caption{The odd spin spectrum arising from the $[1,0]$ of $SO(5)$. Each entry corresponds to the KK tower of $SO(6)$ representations \eqref{KK10}.}
\label{SO5-10-odd}
\end{table}

\section{Partial wave expansion in flat space}
\label{sec:partial_wave_flat}

Next we discuss how the spectrum above contributes to the concrete example the Virasoro-Shapiro amplitude
\beq
\delta^{16}(Q)\frac{ \Gamma \left(- S\right) \Gamma \left(-T\right) \Gamma \left(- U \right) }{\Gamma \left(S +1\right) \Gamma \left( T+1\right) \Gamma \left( U +1\right) } \,,
\label{fullamp}
\eeq
where we use the dimensionless Mandelstam variables
\beq
\begin{gathered}
S = - \frac{\a'}{4} (k_1+k_2)^2\,, \qquad
T = - \frac{\a'}{4} (k_1+k_3)^2\,, \qquad
U = - \frac{\a'}{4} (k_1+k_4)^2\,.
\label{mandelstams}
\end{gathered}
\eeq
The overall delta function $\delta^{16}(Q)$ contains the polarisation dependence, written in an on-shell superspace formalism, and is fixed by supersymmetry. We can directly consider the contribution of exchanged superconformal primaries (as opposed to their full long multiplet) by stripping off this factor. We thus study the function
\beq
A(S,T) = \frac{ \Gamma \left(- S\right) \Gamma \left(-T\right) \Gamma \left(- U \right) }{\Gamma \left(S +1\right) \Gamma \left( T+1\right) \Gamma \left( U +1\right) }\,.
\label{flat_amplitude}
\eeq
This amplitude has poles at $S=\delta$ corresponding to the exchange of states of mass $m^2=4\delta/\alpha'$. Let us briefly discuss how to decompose the residues at these poles into irreducible representations of $SO(4)\times SO(5)$ as listed in \eqref{branching}.
We begin with the usual construction of $SO(9)$ partial waves.
The exchange diagram for a particle at mass level $\delta$ and with $SO(9)$ spin $L$ has the form
	\beq
	\diagramEnvelope{\begin{tikzpicture}[anchor=base,baseline]
		\node (vertL) at (-1,0) [twopt] {};
		\node (vertR) at ( 1,0) [twopt] {};
		\node (opO1) at (-1.4,-1.6) [] {};
		\node (opO2) at (-1.4, 1.6) [] {};
		\node (opO3) at ( 1.4, 1.6) [] {};
		\node (opO4) at ( 1.4,-1.6) [] {};
		\node at (-1.6,-0.8) {$k_1$};
		\node at (-1.6,0.8) {$k_2$};
		\node at (1.6,0.8) {$k_3$};
		\node at (1.6,-0.8) {$k_4$};
		\node at (0,-0.5) {$[L,0,0,0]$};
		\draw [spinning] (opO1) -- (vertL);
		\draw [spinning] (opO2)-- (vertL);
		\draw [spinning] (vertL)-- (vertR);
		\draw [spinning] (opO3) -- (vertR);
		\draw [spinning] (opO4)-- (vertR);
	\end{tikzpicture}}
= \sum\limits_{I=1}^{\dim([L,0,0,0])}
\frac{A_3^I(k_1,k_2) A_3^I(k_4,k_3) }{S-\delta}\,,
\label{diagram}
	\eeq
where the numerator is a sum over on-shell three-point amplitudes
\beq
A_3^I(k_1,k_2) =  \xi^I_{M_1\ldots M_L} \prod\limits_{i=1}^L  \sqrt{\frac{\a'}{2}} (k_1-k_2)^{M_i} \,,
\eeq
defined in terms of an orthonormal basis of polarisation tensors
$\xi^I_{M_1\ldots M_L}$ in the representation $[L,0,0,0]$, which are transverse to the momentum of the particle $k_1+k_2$.
For the vector representation these polarisations satisfy the completeness relation
\beq
\sum\limits_{I=1}^9 \xi_M^I \xi_N^I = \eta_{MN} - \frac{(k_1+k_2)_M (k_1+k_2)_N}{(k_1+k_2)^2}\,,
\eeq
and the completeness relation for the spin $L$ representation can be written in terms of the same vector polarisations (see \cite{Boels:2014dka})
\beq
\sum\limits_{I} \xi_{M_1 \ldots M_L}^I \xi_{N_1 \ldots N_L}^I = 
\pi^{(9)}_{I_1 \ldots I_L,J_1,\ldots J_L}
\prod\limits_{i=1}^L  \xi_{M_i}^{I_i} \xi_{N_i}^{J_i} \,.
\label{completeness_spin}
\eeq
Here we made use of the projector to traceless symmetric rank $L$ tensors of $SO(d)$, which can be most conveniently written by contracting it with vectors $x, \xb \in \mathbb{R}^d$ (see \cite{Costa:2016hju} for details) in terms of a Gegenbauer polynomial
\beq
\pi^{(d)}_{\ell}(x,\xb)
\equiv
x^{I_1}\ldots x^{I_{\ell}}\pi^{(d)}_{I_1 \ldots I_\ell,J_1 \ldots J_\ell} \xb^{J_1}\ldots \xb^{J_{\ell}}
= \frac{\ell!}{2^\ell (\frac{d}{2}-1)_\ell} \left(x^2 \xb^2 \right)^\frac{\ell}{2} C^{(\frac{d}{2}-1)}_\ell \left( \frac{x\cdot \xb}{\sqrt{x^2 \xb^2}} \right)\,.
\label{projector}
\eeq
By combining \eqref{diagram} and \eqref{completeness_spin} one concludes that the residues of the amplitude \eqref{flat_amplitude} have an expansion in terms of the partial waves
\bea
P^{(d)}_\ell (S,T)
={}& \pi^{(d)}_{I_1 \ldots I_\ell,J_1 \ldots J_\ell}
\prod\limits_{i=1}^\ell \frac{\a'}{4 S} \xi_{M_i}^{I_i} (k_1-k_2)^{M_i}  \xi_{N_i}^{J_i} (k_4-k_3)^{N_i}\,,\\
={}&\frac{\ell!}{2^\ell (\frac{d}{2}-1)_\ell} C^{(\frac{d}{2}-1)}_\ell \left(1+ \frac{2T}{S} \right)\,,
\eea{flat_partial_wave}
given by
\beq
\Res\limits_{S=\delta} A(S,T) =  \sum\limits_{L=0}^{2(\delta-1)} A_{\delta,L} P^{(9)}_L(\delta,T)\,, \qquad \delta = 1,2,\ldots\,.
\label{partial_wave_decomposition_10d}
\eeq

In order to find the decomposition into $SO(4) \times SO(5)$ partial waves, we have to implement branching rules as the ones used above in \eqref{branching} in terms of projectors. Labelling an irreducible representation of $SO(d)$ by $\rho_d$, they generally take the form
\beq
\rho_9 = \sum\limits_{[\rho_4;\rho_5] \in \rho_9} N_{[\rho_4;\rho_5]} [\rho_4;\rho_5]\,,
\eeq
where $N_{[\rho_4;\rho_5]}$ are the multiplicities of each representation.
In terms of projectors, this relation takes the form\footnote{This relation also exists for more general representations than traceless symmetric tensors.}
\beq
\pi^{\rho_9}_{\mathbf{I},\mathbf{J}} =
\sum\limits_{[\rho_4;\rho_5] \in \rho_9}
\sum\limits_{k=1}^{N_{[\rho_4;\rho_5]}}
b^{\mathbf{i},\mathbf{m}}_{\mathbf{I}, \rho_9 \to [\rho_4;\rho_5],k} \,
\pi^{\rho_4}_{\mathbf{i},\mathbf{j}}\,
\pi^{\rho_5}_{\mathbf{m},\mathbf{n}}\,
b^{\mathbf{j},\mathbf{n}}_{\mathbf{J}, \rho_9 \to [\rho_4;\rho_5],k}\,.
\label{branching_projectors}
\eeq
Here bold indices encode multiple indices ($I=1,\ldots,9$, $i=1,\ldots,4$, $m=5,\ldots,9$)
\beq
\pi^{\rho_9}_{\mathbf{I},\mathbf{J}} \equiv \pi^{(9)}_{I_1\ldots , J_1 \ldots}\,,\qquad
 \pi^{\rho_4}_{\mathbf{i},\mathbf{j}} \equiv \pi^{(4)}_{i_1\ldots , j_1 \ldots}\,,\qquad
\pi^{\rho_5}_{\mathbf{m},\mathbf{n}} \equiv \pi^{(5)}_{m_1\ldots , n_1 \ldots}\,,
\eeq
and $b^{\mathbf{i},\mathbf{m}}_{\mathbf{I}, \rho_9 \to [\rho_4;\rho_5],k}$ are fixed by the equation \eqref{branching_projectors} itself.
For example, the branching rule
\beq
[1,0,0,0] = [1,1;0,0]+[0,0;1,0]\,,
\eeq
is written in terms of the projectors for vectors $\pi^{(d)}_{I,J} = \delta_{I,J}$ and singlets $\pi^{(d)} = 1$ as
\beq
\pi^{[1,0,0,0]}_{I,J} = \delta_I^i \pi^{[1,1]}_{i,j} \pi^{[0,0]} \delta_J^j
+ \delta_I^m \pi^{[0,0]} \pi^{[1,0]}_{m,n} \delta_J^n = \delta_{I,J}\,.
\eeq
In order to find the partial waves for $SO(4) \times SO(5)$, we simply have to insert \eqref{branching_projectors} into \eqref{flat_partial_wave}.
In the case where all momenta along the $SO(5)$ directions vanish $k_{1,2,3,4}^m=0$ we see that only the singlet of $SO(5)$ can be exchanged.
In this case there is a relation between $SO(9)$ and $SO(4)$ partial waves
\beq
P^{(9)}_L (S,T)\bigg|_{k_{1,2,3,4}^m=0} = \sum\limits_{k=0}^{\lfloor \frac{L}{2} \rfloor}
B_{S,k} P^{(4)}_{L-2k} (S,T)\,,
\eeq
and \eqref{partial_wave_decomposition_10d} becomes a decomposition into 5D Lorentzian partial waves\footnote{The coefficients $a_{\delta,\ell}$ are related to the coefficients  $f_0(\delta,\ell)$ introduced in \cite{Alday:2022uxp} by
\beq
a_{\delta,\ell} = \frac{2^\ell}{\delta^2 (\ell+1)} f_0(\delta,\ell)\,.
\eeq}
\beq
\Res\limits_{S=\delta} A(S,T) =  \sum\limits_{\ell=0}^{2(\delta-1)} a_{\delta,\ell} P^{(4)}_\ell(\delta,T)\,, \qquad \delta = 1,2,\ldots\,.
\label{partial_wave_decomposition_flat}
\eeq
Using the explicit form for the partial waves for $d=4$ 
\begin{equation}
P^{(4)}_\ell(S,T)=\frac{\ell+1}{2^\ell} \, _2F_1\left(-\ell,\ell+2;\frac{3}{2};-\frac{T}{S}\right)\,,
\end{equation}
and the explicit form of the residues at $S=\delta$ we obtain

\beq
\frac{(T+1)_{\delta-1}^2}{\Gamma(\delta+1)^2}  = \sum_{\ell=0}^{2(\delta-1)} a_{\delta,\ell} \frac{\ell+1}{2^\ell} \, _2F_1\left(-\ell,\ell+2;\frac{3}{2};-\frac{T}{\delta}\right)\,,
\label{partial_wave_decomposition_flat_explicit}
\eeq
which allows to solve for the coefficients $a_{\delta,\ell}$ for $\delta=1,2,\cdots$. These coefficients will serve as input to compute $\cN = 4$ SYM structure constants at leading order in $1/\lambda$.

\section{Partial wave expansion in AdS}
\label{sec:partial_wave_ads}
In this section we consider the four-point correlator of arbitrary chiral primary operators in $\cN = 4$ SYM theory at leading non-trivial order in a $1/c$ expansion, where $c=(N^2-1)/4$ is the central charge. This is related to \eqref{fullamp} in the flat space limit. More precisely, consider the four-point function of four scalar superconformal primaries of $1/2$-BPS multiplets $\cO_{p_i}(x_i,y_i)$ with conformal dimension and R-charge $p_i$ where $\sum_{i=1}^4 p_i$ is even
\beq
\< \cO_{p_1}(x_1,y_1) \cdots \cO_{p_4}(x_4,y_4) \>\,.
\eeq
$y_i \in \mathbb{R}^6$ are polarisations satisfying $y_i^2=0$ that we contracted with the $p_i$ R-symmetry indices of each operator.
Similar to the amplitude in flat space, we strip off the part of the correlator that is fixed by supersymmetry and focus on the reduced correlator $\cT(z,\zb,\alpha,\ab)$. The reduced correlator admits an expansion into superconformal long multiplets, labelled by their superconformal primary with dimension $\Delta$, spin $\ell$ and $SO(6)$ representation ${\cal R}$,\footnote{We suppress an overall factor $1/c$.}
\begin{align}
\cT (z,\zb,\al,\ab) ={}&
 \sum\limits_{\De,\ell,{\cal R}} C^{\{p_1 p_2\}}_{\De,\ell,{\cal R}} C^{\{p_3 p_4\}}_{\De,\ell,{\cal R}} G^{r,s}_{\De+4,\ell} (z,\bar{z})Z_{{\cal R}}^{r,s}(\al,\ab)\,.
\label{OPE}
\end{align}
For the precise relation between the full and reduced correlator and the definition of the superconformal blocks, see appendix \ref{mixedCorr}. The cross-ratios are given by
\bea
U= \frac{x_{12}^{2}x_{34}^2}{x_{13}^2x_{24}^2} = z \bar{z}, \qquad &V = \frac{x_{23}^2x_{14}^2}{x_{13}^2x_{24}^2} = (1-z)(1-\bar{z}), \\
\frac{1}{\sigma}= \frac{y_{12} y_{34}}{y_{13} y_{24}} = \alpha \bar{\alpha}, \qquad& \frac{\tau}{\sigma} = \frac{y_{23} y_{14}}{y_{13} y_{24}} = (1-\alpha)(1-\bar{\alpha})\,,
\eea{cross-ratios}
and $r=p_{21}=p_2-p_1$, $s=p_{34}$.
In the flat space limit, the reduced correlator is related to the reduced amplitude \eqref{flat_amplitude}.
This relation is most explicit in terms of the Mellin amplitude $M(s, t,\al,\ab)$, defined through the Mellin transform
\beq
\cT(z,\zb,\al,\ab)=\int_{-i\infty}^{i \infty} \frac{ds  dt }{(4\pi i)^2}U^{s/2+2} V^{\frac{t-p_2-p_3}{2}} \Gamma_{p_1 p_2 p_3 p_4} M(s, t,\al,\ab)\, ,
\label{mellin}
\eeq
where we have introduced Mellin variables $s,t, u$ satisfying $s+t+u=\sum_{i=1}^4 p_i-4$ and
\bea
\Gamma_{p_1 p_2 p_3 p_4}={}&\Gamma \mleft( \frac{p_1+p_2-s}{2}\mright) \Gamma \mleft( \frac{p_3+p_4-s}{2}\mright) \Gamma \mleft( \frac{p_1+p_4-t}{2}\mright) \Gamma \mleft( \frac{p_2+p_3-t}{2}\mright) \\
&\times \Gamma \mleft( \frac{p_1+p_3-u}{2}\mright) \Gamma \mleft( \frac{p_2+p_4-u}{2}\mright)\, .
\eea{gammas}

In the flat space limit we are considering, where $s,t$ become large while $\alpha,\bar \alpha$ are kept fixed, we only access the singlet of $SO(5)$. In this limit the whole Mellin amplitude must factorise into a polynomial $P(\a,\ab)$ which encodes the KK modes for the singlet of $SO(5)$, times a function $\tilde{M}(s,t)$ of the Mellin variables which encodes the $AdS_5$ directions
\beq
\lim\limits_{s,t\to \infty} M(s,t,\a,\ab) = \tilde{M}(s,t) P(\a,\ab) \,.
\label{M_factorisation}
\eeq
The AdS$_5$ factor $\tilde{M}(s,t)$ is related to the flat space amplitude \eqref{flat_amplitude} by the flat space limit
\beq \label{flat}
A(S,T) = - 2 \lambda^\frac{3}{2}  \lim\limits_{R \to \infty} \int_{\kappa-i\infty}^{\kappa+ i \infty} \frac{d\alpha}{2 \pi i} \, e^\alpha \alpha^{-2-\frac12 \sum_i p_i} 
\tilde{M} \left( \frac{2\sqrt{\lambda} S}{\alpha}, \frac{2\sqrt{\lambda} T}{\alpha} \right)  \,.
\eeq
From this we immediately find
\bea
A(S,T) &{}= - \frac{1}{STU} - 2 \sum\limits_{a,b=0}^\infty  \hat\sigma_2^a \hat\sigma_3^b    \alpha^{(0)}_{a,b}\,,\\
\quad \Rightarrow \quad
\tilde{M}(s,t) &{}= \frac{4 \Gamma(\frac12 \sum_i p_i -1)}{(s-2)(t-2)(u-2)} + \sum\limits_{a,b=0}^\infty  \frac{\Gamma(2a+3b+\frac12 \sum_i p_i + 2)}{8^{a+b}\lambda^{\frac32 + a + \frac32 b} } \sigma_2^a \sigma_3^b \alpha^{(0)}_{a,b}\,,
\eea{A_to_M}
where we use the bases of crossing-symmetric polynomials
\bea
\hat\sigma_2 &= \frac12 ( S^2+T^2+U^2)\,, \qquad
\hat\sigma_3 = STU\,, \qquad
\sigma_2 &= s^2+t^2+u^2 \,, \qquad
\sigma_3 = s t u\,.
\eea{sigma}

To determine the factor $P(\a,\ab)$ we can make use of the known supergravity term in the Mellin amplitude. 
Taking the large $s,t$ limit of the result from \cite{Rastelli:2016nze,Aprile:2018efk} (assuming without loss of generality $p_{21} \geq 0$, $p_{43}\geq 0$ and $p_{43}\geq p_{21}$) and comparing to \eqref{M_factorisation} we find that
\beq
\lim\limits_{s,t\to \infty} M(s,t,\a,\ab) = \frac{4 \Gamma(\frac12 \sum_i p_i -1)}{s t u}
P(\a,\ab)+ O(\lambda^{-3/2}) \,, 
\label{sugra_fsl}
\eeq
with
\begin{align}
P(\a,\ab) &= \frac{\sqrt{p_1 p_2 p_3 p_4} }{2 \Gamma(\frac12 \sum_i p_i -1)} \sum\limits_{\substack{0\leq i,j,k\leq \mathcal{E}-2\\i+j+k=\mathcal{E}-2}} \frac{\sigma^{\frac{p_{43}}{2} + i} \tau^j}{i! j! k! (i + \frac{p_{43}+p_{21}}{2})! (j+ \frac{p_{43}-p_{21}}{2})! (k+|\frac{p_1+p_2-p_3-p_4}{2}|)!}\,,\nonumber\\
\mathcal{E} &= p_3 - \min(0, \tfrac{p_1+p_2-p_3-p_4}{2})\,,
\label{P}
\end{align}
see also \cite{Alday:2020dtb}. Next we want to extract from this the OPE coefficients in the large $\lambda$ limit of the expansion \eqref{OPE}. For the AdS factor this can be done by taking the flat space limit of the conformal partial wave decomposition
\beq
\tilde{M}(s,t)=\sum_{\ell=0}^{\infty}\int_{-i\infty}^{i\infty} d\nu \, b_\ell(\nu^2)M_{\nu,\ell}(s,t) \,,
\eeq
where $M_{\nu,\ell}(s,t)$ is the Mellin transform of a conformal partial wave  as defined in \cite{Costa:2012cb} and $b_\ell(\nu^2)$ has poles at the conformal dimensions of superconformal primaries with residues
\begin{align}
b_\ell(\nu^2)\approx b_{\Delta,\ell} \frac{2^\ell K^{\{p_i+2\},4}_{\Delta+4,\ell}}{\nu^2+(\Delta+4-2)^2}\,, 
\end{align}
where
\bea
K^{\{\Delta_i\},d}_{\Delta,\ell} ={}& \frac{\Gamma(\Delta+\ell) \,\Gamma(\Delta-\frac{d}{2}+1) \, (\Delta-1)_\ell   }{ 
4^{\ell-1} 
\Gamma\!\left( \frac{\Delta +\ell+\Delta_{1 2}}{2}\right)
\Gamma\!\left( \frac{\Delta +\ell-\Delta_{1 2}}{2}\right) 
\Gamma\!\left( \frac{\Delta +\ell+\Delta_{3 4}}{2}\right)
\Gamma\!\left( \frac{\Delta +\ell-\Delta_{3 4}}{2}\right)}\\
&\frac{1}{
 \Gamma\!\left( \frac{\Delta_1 +\Delta_{2} -\De+\ell}{2}\right)
\Gamma\!\left( \frac{\Delta_3 +\Delta_{4} -\De+\ell}{2}\right)
\Gamma\!\left( \frac{\Delta_1 +\Delta_{2} +\De+\ell-d}{2}\right)
\Gamma\!\left( \frac{\Delta_3 +\Delta_{4} +\De+\ell-d}{2}\right)}\,.
\eea{KDeltaJ}
As shown in \cite{Costa:2012cb,Goncalves:2014ffa,Alday:2022uxp}, we can use
\beq
K^{\{p_1+2,p_2+2,p_3+2,p_4+2\},4}_{\Delta+4,\ell} \underset{\Delta\gg1}{\approx}   \left(\frac{\Delta}{2} \right)^{8-\sum_{i} p_i}
\frac{2^{2 \Delta +2 \ell+17}  }{\pi ^3 \Delta ^{2 (\ell+5)}}
\sin\left(\tfrac{\pi(\Delta-\ell-p_1 -p_2)}{2}     \right)
\sin\left(\tfrac{\pi(\Delta-\ell-p_3 -p_4)}{2}     \right)\,,
\eeq
together with the fact that the flat space limit of $M_{\nu,\ell}(s,t)$ is the 5D flat space partial wave which appears in  \eqref{partial_wave_decomposition_flat} to show that
\beq
b_{\Delta,\ell} = \left(\frac{\Delta}{2} \right)^{\sum_{i} p_i-8} \frac{ \pi^3 \Delta^{10}}{\lambda  2^{2\Delta+\ell+16} \sin(\frac{\pi}{2} ( \Delta-\ell-p_1 -p_2)) \sin(\frac{\pi}{2} ( \Delta-\ell-p_3 -p_4))} a_{\delta,\ell}\,.
\label{b}
\eeq
The $S^5$ factor simply needs to be expanded into spherical harmonics
\beq
P(\a,\ab) = \sum\limits_{n=\max(p_{21},p_{43})}^{\min(p_1+p_2,p_3+p_4)-4} h_{n} Z_{[n,0,0]}(\a,\ab)\,,
\label{P_expansion}
\eeq
where $n$ increases in steps of 2 and the coefficients can be written in terms of sphere overlap integrals defined in \eqref{S_def} below
\beq
h_n = \frac{S_{p_1-2,p_2-2,n}S_{p_3-2,p_4-2,n}}{
\pi^6 \prod_{i=1}^4 \sqrt{ \Gamma(p_i)\Gamma(p_i-1)}}\,.
\label{h}
\eeq
In this way we find the leading OPE coefficients
\beq
C^{\{p_1 p_2\}}_{\De,\ell,[n,0,0]} C^{\{p_3 p_4\}}_{\De,\ell,[n,0,0]} =
b_{\Delta,\ell} h_{n} 
= \lambda_{p_1,p_2,n} \lambda_{p_3,p_4,n} \frac{ \pi^3 \Delta^{10}}{\lambda  2^{2\Delta+\ell+16} } a_{\delta,\ell}\,,
\label{C_result}
\eeq
where we defined
\beq
\lambda_{p_1,p_2,n} =  \left(\frac{\Delta}{2} \right)^{p_1+p_2-4}
\frac{1}{\sin(\frac{\pi}{2} ( \Delta-\ell-p_1 -p_2))}
\frac{S_{p_1-2,p_2-2,n}}{
\pi^3 \sqrt{ \Gamma(p_1)\Gamma(p_1-1) \Gamma(p_2)\Gamma(p_2-1)}} \,.
\eeq
Let us make some comments on this result.
The full supergravity correlator of \cite{Rastelli:2016nze} that we used to compute the limit \eqref{sugra_fsl} decomposes into all $SO(6)$ representations of the form $[n-m,m,m]$, however \eqref{P_expansion} shows that only the symmetric tensors $[n,0,0]$ survive the flat space limit. This is consistent with the fact that this limit explores only the singlet of $SO(5)$. Moreover, \eqref{h} implies that $P(\a,\ab)$ is the sum over products of two sphere overlap integrals.\footnote{Up to factors $\sqrt{ \Gamma(p_i)\Gamma(p_i-1)}$ which can be attributed to the normalisation of the external operators.} We state this more explicitly in \eqref{P_vs_overlaps} below.

\subsection{Operator mixing}
\label{sec:mixing}

We can now turn to the question on whether studying these general correlators allows us to unmix the OPE data for heavy operators extracted from these correlators.

The result \eqref{C_result} should really be seen as a sum over products of OPE coefficients for multiple degenerate operators. Let us introduce a label $I$ for these species and write more precisely
\beq\label{GeneralSumOPE}
C^{\{p_1 p_2\}}_{\De,\ell,[n,0,0]} C^{\{p_3 p_4\}}_{\De,\ell,[n,0,0]} = \sum\limits_{I=1}^{N(\Delta, \ell, n)} C^{\{p_1 p_2\} I}_{\De,\ell,[n,0,0]} C^{\{p_3 p_4\} I}_{\De,\ell,[n,0,0]} \,,
\eeq
where $N(\Delta, \ell, n)$ is the number of superprimaries with the same quantum numbers $(\Delta, \ell, n)$ at strong coupling.
The quantity (\ref{GeneralSumOPE}) is what we actually computed. Let us explain what equation \eqref{C_result} implies for the individual $C^{\{p_1 p_2\} I}_{\De,\ell,[n,0,0]}$.

Let us introduce the following vectors 
\beq
A^I = C^{\{p_1 p_2\} I}_{\De,\ell,[n,0,0]}\,, \qquad
B^I = C^{\{p_3 p_4\} I}_{\De,\ell,[n,0,0]}\,.
\eeq
The triangle inequality for these two vectors reads $(A \cdot B) ^2 \leq |A|^2 |B|^2$. However, equation \eqref{C_result} implies that actually 
\beq
(A \cdot B )^2  = |A|^2 |B|^2\,.
\eeq
This implies that $A^I$ and $B^I$ are parallel vectors. Thus we conclude that 
\beq
C^{\{p_1 p_2\} I}_{\De,\ell,[n,0,0]} = \chi^{\{p_1 p_2\}}_{\De,\ell,[n,0,0]} D^I\,,
\eeq
where 
\beq
\chi^{\{p_1 p_2\}}_{\De,\ell,[n,0,0]}= \pm \lambda_{p_1,p_2,n} \sqrt{ \frac{ \pi^3 \Delta^{10}}{\lambda  2^{2\Delta+\ell+16} } a_{\delta,\ell} }\,,
\eeq 
and $D^I$ is a unit vector $|D|^2 = 1$. It is not possible to say more about the individual components  $D^I$, i.e.\ we cannot determine them at leading order in a way similar to \cite{Alday:2017xua,Aprile:2017bgs,Aprile:2017xsp} for double trace operators. The reason for this is ultimately that all the KK modes originate from the same amplitude in 10-dimensional flat space.
We expect that it would be possible to unmix the OPE coefficients by considering correlators with different flat space analogues, for example by taking some of the external operators to be massive as well.

\section{Comparison with previous results}
\label{sec:comparison}

In this section we compare our results with those of \cite{Minahan:2014usa}, where the three-point functions for two chiral primaries and one massive string operator on the leading Regge trajectory were computed from the flat space limit.

The method of \cite{Minahan:2014usa} requires that the dimensions of all three operators in the three-point function ($p_1$, $p_2$ and $\Delta$) are large, of order $\lambda^{1/4}$.
On first sight this limit seems to be incompatible with our assumption that $p_1,p_2$ are much smaller than $\lambda^{1/4}$ in the large $\lambda$ limit. This assumption led to the conclusion that to leading order in $1/\lambda$ only the 
KK modes of the singlet of $SO(5)$ couple to two chiral primaries, see section \ref{sec:spectrum_correlator}. The reason that we can compare results anyway is that,
as mentioned in section \ref{sec:so5_vector}, the leading Regge trajectory arises only from the singlet of $SO(5)$ in general, so we can lift the restriction $p_i \ll \lambda^{1/4}$
when considering the structure constants of these operators.
In order to compare, it is thus enough to take the additional limit $\Delta \gg p_1,p_2$ of the results of \cite{Minahan:2014usa}.

The structure constants given in \cite{Minahan:2014usa} are those for the maximal superconformal descendants of the operators we considered above. These operators are conformal, but not superconformal, primaries. For the external operators these are given by
\beq
\cL_p = L^+_p + L^-_p = Q^4 \cO_p + \bar{Q}^4 \cO_p\,,
\label{Lrel}
\eeq
and they are scalars with conformal dimension $p+2$ in the $[p-2,0,0]$ of $SO(6)$.
For $p=2$ this operator is the Lagrangian of the theory.
We consider the decomposition of the crossing symmetric correlator into bosonic conformal blocks
 \begin{align}
&\left\langle \mathcal{L}_p(x_1) \mathcal{L}_p(x_2) \mathcal{L}_p(x_3) \mathcal{L}_p(x_4)  \right\rangle 
=\frac{G(z,\zb,\a,\ab)}{x_{12}^{4p}x_{34}^{4p}} = \frac{1}{x_{12}^{4p}x_{34}^{4p}} \sum_{\Delta,\ell,n}\tl C^2_{\Delta,\ell,n}G_{\Delta,\ell}(z,\zb)Z_{[n,0,0]}(\a,\ab)  \,.
\label{L_ope}
\end{align}
The OPE coefficients $\tl C_{\Delta,\ell,n}$ were computed in \cite{Minahan:2014usa} for operators on the leading Regge trajectory, which are
descendants of the massive short superconformal primaries discussed above $Q^4 \bar{Q}^4 \cO_{\delta,\ell,[n,0,0]}$ and have dimension $\Delta+4$, spin $\ell+4$ and R-charge $n$.
In terms of the quantum numbers
\beq
\De_{1,2} = p_{1,2}+2\,, \quad
n_{1,2} = p_{1,2}-2\,, \quad
\De_3 = 2 \sqrt{\frac{\ell+2}{2}} \lambda^\frac{1}{4} + 2 + O(\lambda^\frac{1}{4})\,, \quad S = \ell + 4\,, \quad n_3 = n\,, 
\eeq
their result is\footnote{The factor $2^{\frac{3-S}{2}}$ accounts for our different normalisation of conformal blocks and the whole correlator compared to \cite{Minahan:2014usa,Costa:2012cb,Goncalves:2014ffa}.}
\beq
\tl C_{\Delta+4,\ell+4,n} = 2^{\frac{3-S}{2}} \frac{\sqrt{(\Delta_1-1)(\Delta_2-1)(\Delta_3-1)}}{2^{5/2}\pi}\,\frac{\Gamma(\al_1)\Gamma(\al_2)\Gamma(\al_3)\Gamma(\Sigma-2)}{\Gamma(\Delta_1)\Gamma(\Delta_2)\Gamma(\Delta_3)}\,\cG_{123}\,,
\label{Ctilde}
\eeq
with
\beq
\Sigma=\half(\Delta_1+\Delta_2+\Delta_3) \,, \qquad\alpha_i = \Sigma - \Delta_i \,,
\eeq
and the coupling
\beq
\cG_{123}=
\frac{8 N^2 \sqrt{\lambda}}{\pi^2}\,\<V_{\cL} V_{\cL} V_{S}\> S_{n_1,n_2,n_3}\,.
\label{coupling}
\eeq
The factor $\<V_{\cL} V_{\cL} V_{S}\>$ was computed using flat space vertex operators as
\beq
\<V_{\cL} V_{\cL} V_{S}\>= \frac{(2\a'\alpha_1 \alpha_2 \alpha_3 \Sigma)^{S/2}}{N^3(\Delta_3)^S\Gamma(S/2)} \,,
\eeq
and the factor $S_{n_1,n_2,n_3}$ in \eqref{coupling} comes from a sphere overlap integral in $S^5$ and will be discussed in appendix \ref{sec:sphere_integrals} below.

The correlator \eqref{L_ope} is related by a differential operator to the one of superconformal primaries. In \cite{Alday:2022uxp} we used the differential operator for the case $\< 2222 \>$ given in \cite{Drummond:2006by,Goncalves:2014ffa} to relate the OPE coefficients for the two correlators. Here we will use the general differential operator given in \cite{Caron-Huot:2018kta} to generalise this computation to the case $\< pppp \>$.

In \cite{Caron-Huot:2018kta} the authors provided the differential operator
\beq
 \Delta^{(8)} =
 \frac{z \zb\a\ab}{(z-\zb)(\a-\ab)} \left( \cD_z -\cD_\a\right)\left( \cD_z -\cD_\ab\right)\left( \cD_{\zb} -\cD_\a\right)\left( \cD_{\zb} -\cD_{\ab}\right)
 \frac{(z-\zb)(\a-\ab)}{z \zb\a\ab}\,,
\label{Delta8}
\eeq
where
\beq
\cD_x \equiv x^2\partial_x(1-x)\partial_x -\tfrac12(r+s)x^2\partial_x -\tfrac14 rs x \,,
\label{Casimir}
\eeq
which acts on the reduced correlator of $\cO_p$ introduced in \eqref{OPE} and computes
\beq
\< L^+_p (x_1) L^+_p (x_2) L^-_p (x_3) L^-_p (x_4)  \> \equiv \frac{\tl G(z,\zb,\a,\ab)}{x_{12}^{4p}x_{34}^{4p}} = \frac{2}{x_{12}^{4p}x_{34}^{4p}} \frac{\Delta^{(8)}}{p^4 (p^2-1)^2}  \cT(z,\zb,\a,\ab)\,.
\eeq
We then use the relation \eqref{Lrel} to write
\begin{align}
{}& \< \cL_p \cL_p \cL_p \cL_p  \> = 2 \left( \< L^+_p L^+_p L^-_p L^-_p  \> + \< L^-_p L^+_p L^+_p L^-_p  \> + \< L^+_p L^-_p L^+_p L^-_p  \> \right)\nonumber\\
={}& \frac{2}{x_{12}^{4p}x_{34}^{4p}} \bigg( \tl G(z,\zb,\a,\ab) 
+ \left( \tfrac{z \zb}{(1-z)(1-\zb)} \right)^{p+2} \left( \tfrac{(1-\a)(1-\ab)}{\a \ab} \right)^{p-2} \tl G(1-z,1-\zb,1-\a,1-\ab) \nonumber\\
&+  \frac{(z \zb)^{p+2}}{(\a \ab)^{p-2}} \tl G(1/z,1/\zb,1/\a,1/\ab)   \bigg)\,.
\label{L_use_crossing}
\end{align}
Using also the crossing symmetry of $\cT(z,\zb,\a,\ab)$, this relation provides us with a differential operator $\tl\Delta^{(8)}$ relating $\cT(z,\zb,\a,\ab)$ and $G(z,\zb,\a,\ab)$
\beq
G(z,\zb,\a,\ab) = \tl\Delta^{(8)} \, \cT(z,\zb,\a,\ab)\,.
\eeq
Following appendix F of \cite{Alday:2022uxp} we can act on a generic superconformal block contributing to $\cT(z,\zb,\a,\ab)$ and obtain the precise contributions of the bosonic conformal blocks for each component of this supermultiplet contributing to the correlator $G(z,\zb,\a,\ab)$.
In order to compare to the result \eqref{Ctilde} we can make two big simplifications. We only need the contribution of the bottom component of the supermultiplet, which has spin $\ell +4$, because this is the only component that contributes to the leading Regge trajectory in the sense of the expansion \eqref{L_ope}. Furthermore we can take $\Delta$ to be very large as before. The result is
\begin{align}
\label{interesting_L_block}
\cT(z,\zb,\a,\ab) &= G_{\Delta+4,\ell}(z,\zb)Z_{[n,0,0]}(\a,\ab) \Rightarrow \\
G(z,\zb,\a,\ab) &= \frac{\Delta^8+ O(\Delta^7)}{2^{13} p^4 (p^2-1)^2} G_{\Delta+4,\ell+4}(z,\zb)Z_{[n,0,0]}(\a,\ab)  + \text{subleading Regge trajectories}\,.
\end{align}
Using this relation and the result for the flat space partial wave coefficients \cite{Alday:2022uxp}
\beq
a_{\frac{\ell}{2}+1,\ell} = \frac{(\ell+2)^{\ell-2}}{4^{\ell-1}\Gamma(\frac{\ell}{2}+1)^2}\,,
\eeq
we can now check that for operators on the leading Regge trajectory, our result \eqref{C_result} agrees with \eqref{Ctilde} once we expand at large $\Delta$
\beq
\frac{\Delta^4 +O(\Delta^{\frac72})}{2^{\frac{13}{2}} p_1 p_2 \sqrt{ (p_1^2-1)(p_2^2-1)}} C^{\{p_1 p_2\}}_{\De,\ell,[n,0,0]}
= \tl C_{\Delta+4,\ell+4,n}\,.
\eeq

\section{Conclusions}
\label{sec:conclusion}

In this paper we studied the spectrum and structure constants of massive short operators in planar $\cN = 4$ SYM theory at strong coupling by considering strings on $AdS_5 \times S^5$ in the flat space limit. It would be very interesting to compare the spectrum of degeneracies with explicit results from integrability. Where data from integrability is available, on the leading Regge trajectory, the degeneracies do match. More generally, we cannot rule out the possibility that two different operators at finite $\lambda$ become identical in the strict $\lambda=\infty$ limit, and are seen as a single operator in flat space. It would be interesting to explore this. 

It would also be very interesting to reproduce our results for structure constants, from integrability. Note in particular that our results are valid for large $R-$charges (as long as they are smaller than $\lambda^{1/4}$), where integrability computations are expected to simplify.  See \cite{Basso:2022nny} for relevant progress in this direction. 

Our result for the structure constants of two chiral primaries and one massive short operator implies that the OPE data contained in the AdS Virasoro-Shapiro amplitude of \cite{Alday:2022uxp,Alday:2022xwz} cannot be unmixed at leading order by considering generalisations to correlators of arbitrary chiral primary operators. Possible alternatives are to unmix the OPE data at a higher order in the $1/\sqrt{\lambda}$ expansion or to consider other correlators, for example ones with external massive operators.

Of course it is possible to apply the methods of \cite{Alday:2022uxp,Alday:2022xwz} to more general correlators to compute corrections to the OPE data of massive operators with R-charge. We plan to report on this in a forthcoming paper.

Another possible avenue is to implement a flat space limit that probes large R-charges of the external operators $p_i \sim \lambda^{\frac14}$ and accesses the full spectrum of strings in 10 dimensions.

\section*{Acknowledgements} 

The work of LFA and TH is supported by the European Research Council (ERC) under the European Union's Horizon
2020 research and innovation programme (grant agreement No 787185). LFA is also supported in part by the STFC grant ST/T000864/1. JS is supported by the STFC grant ST/T000864/1. 

\appendix

\section{Correlator Setup}
\label{mixedCorr}

In this appendix we state the definitions for the reduced correlator and long superconformal block following the conventions of \cite{Caron-Huot:2018kta}.

\subsection{Reduced correlator}

We normalise the two point function of $\frac12$-BPS operators of dimension $p$ as
\bea
\< \cO_{p}(x_1,y_1) \cO_{p}(x_2,y_2) \> = \frac{(y_1 \cdot y_2)^{p}}{|x_1-x_2|^{2p}}\,,
\eea{Eq:Normalization}
and write their four-point functions as
\bea
\< \cO_{p_1}(x_1,y_1) \cdots \cO_{p_4}(x_4,y_4) \>
&\equiv 
T_{p_1 p_2 p_3 p_4}
\mathcal{G}(z,\zb,\al,\ab)\,,\\
T_{p_1 p_2 p_3 p_4} &= g_{12}^{\frac{p_1+p_2}{2}}
g_{34}^{\frac{p_3+p_4}{2}}
\left(\frac{g_{24}}{g_{14}}\right)^{\frac{p_{2}-p_1}{2}}
\left(\frac{g_{13}}{g_{14}}\right)^{\frac{p_{3}-p_4}{2}}\,,
\eea{Eq:OriginalCorrelator}
where $g_{ij} = \frac{y_i \cdot y_j}{x_{ij}^2}$.
The superconformal Ward identities imply that the correlator can generally be written in the form \cite{Dolan:2004iy}
\begin{align}
\mathcal{G}(z, \zb, \al, \ab) &= k \chi(z, \al) \chi(\zb, \ab) + \frac{(z-\al)(z-\ab)(\zb-\al) (\zb - \ab)  }{(\al - \ab)(z - \zb)} \nonumber \\
&\times \left(- \frac{ \chi(\zb,\ab) f(z,\al)}{\al z (\zb-\ab)} + \frac{\chi(\zb,\al) f(z,\ab)}{\ab z (\zb-\al)} + \frac{\chi(z,\ab) f(\zb,\al)}{\al \zb (z-\ab)} - \frac{\chi(z,\al) f(\zb,\ab)}{\ab \zb (z-\al)} \right) \nonumber \\
&+ \frac{(z-\al) (z - \ab) (\zb- \al) (\zb - \ab)}{(\al \ab)^2 (z \zb)^2} H(z,\zb, \al, \ab), \label{G ansatz}
\end{align}
where $k$ is called the unit contribution, $f$ is the chiral correlator and $H$ is the reduced correlator.
These different parts can be extracted from $\mathcal{G}$ as follows
\be\label{kf from G} \begin{aligned}
 k &= \cG(z, \zb, z, \zb), \\
 f(\zb,\ab) &= \frac{\ab \zb}{\zb - \ab} \left( \cG(z, \zb, z, \ab) - k \chi(\zb,\ab)  \right),
\end{aligned}\ee
and $H$ is obtained by subtracting the other contributions from \eqref{G ansatz}. The function $\chi$ is given by
\begin{equation}
\chi(z,\al) = \left( \frac{z}{\alpha} \right)^{\max(|p_{21}|,|p_{34}|)/2} \left( \frac{1-\al}{1-z} \right)^{\max(p_{21}+p_{34},0)/2}.
\label{Eq:ChiWard}
\end{equation}
We further define the interacting part of the reduced correlator by subtracting the reduced correlator of the free theory
\beq
\cT(z,\zb, \al, \ab) = H(z,\zb, \al, \ab) - H^\text{free}(z,\zb, \al, \ab)\,.
\eeq

\subsection{Superconformal blocks}
\label{sec:superconformal_blocks}

The superconformal multiplets that can be exchanged in the four-point functions we will consider can be labelled by the conformal dimension $\De$, spin $\ell$ and SO(6) representation $[n-m,m,m]$ of the superconformal primary of the multiplet.

Massive string states are in long multiplets, which have the following contributions to the unit contribution, chiral correlator and reduced correlator
\bea
\mathcal{A}_{\De,\ell,m,n}^{r,s} = \left\{\begin{array}{l}
\displaystyle
k=0, \qquad
f(z,\al) = 0,
\\
\displaystyle 
H(z,\zb,\al,\ab) =  G^{r,s}_{\De+4,\ell} (z,\bar{z})Z_{[n-m,m,m]}^{r,s}(\al,\ab)
\end{array}\right..
\eea{A multiplet}
These definitions are in terms of
the usual $SO(4,2)$ conformal blocks
\begin{align}
G_{\Delta,\ell}^{r,s} (z,\bar{z}) &= \frac{z \bar{z}}{\bar{z}-z} \left[ k_{\frac{\Delta-\ell-2}{2}}^{r,s}(z) k_{\frac{\Delta+\ell}{2}}^{r,s}(\bar{z}) - k_{\frac{\Delta+\ell}{2}}^{r,s}(z) k_{\frac{\Delta-\ell-2}{2}}^{r,s}(\bar{z}) \right], \label{Eq:G-block}
\\
k_h^{r,s}(z) &= z^h \; _{2}F_1 \left(h+ \frac{r}{2}, h + \frac{s}{2}; 2h, z \right),
\end{align}
and the $S_5$ spherical harmonics, which are given in terms of the same functions
\beq
Z_{[n-m,m,m]}^{r,s} (\al,\ab) = (-1)^{m} G_{-n,m}^{-r,-s}(\al,\ab)\,.
\label{eq:Z}
\eeq

\section{Sphere integrals}
\label{sec:sphere_integrals}

Let us now relate the sphere overlap factor in the OPE coefficients to the four-point partial wave \eqref{eq:Z}.
To encode the irreducible representation $[n,0,0]$ of $SO(6)$ we introducing an orthonormal basis of traceless symmetric $SO(6)$ tensors $C^I_{i_1 \ldots i_n}$ labelled by $I=1,\ldots,\dim ([n,0,0])$ which satisfy
\beq
C^{I_1}_{i_1 \ldots i_n} C^{I_2}_{i_1 \ldots i_n} = \delta^{I_1 I_2}\,,
\eeq
as well as the completeness relation
\beq
C^{I}_{i_1 \ldots i_n} C^{I}_{j_1 \ldots j_n} = \pi^{(6)}_{i_1 \ldots i_n,j_1 \ldots j_n}\,,
\label{completeness}
\eeq
where $\pi^{(6)}_{i_1 \ldots i_n,j_1 \ldots j_n}$ is the projector to traceless symmetric $SO(6)$ representations \eqref{projector}. The scalar spherical harmonics are defined by
\beq
Y^I_n = z(n) C^{I}_{i_1 \ldots i_n} \xi^{i_1} \ldots \xi^{i_n}\,,
\eeq
where $\xi \in S^5$, i.e.\ $\xi^2=1$ and the normalisation
\beq
z(n) = \sqrt{2^{n-1} (n+1)(n+2)}\,,
\eeq
is chosen such that
\beq
\int_{S^5} d \xi \, Y^{I_1}_n Y^{I_2}_n = \omega_5 \delta^{I_1 I_2}\,,
\eeq
where $\omega_5 = \pi^3$ is the volume of $S^5$.
To connect to our sphere polarizations $y^i$ (satisfying $y^2=0$) from before we introduce the object
\beq
C^I_n (y) = C^{I}_{i_1 \ldots i_n} y^{i_1} \ldots y^{i_n}\,.
\eeq
The structure constants were given in \cite{Minahan:2014usa} in terms of the sphere overlap integrals
\beq
\< \psi_{n_1} (y_1) \psi_{n_2} (y_2) \psi_{n_3} (y_3) \> = C^{I_1}_{n_1} (y_1) C^{I_2}_{n_2} (y_2) C^{I_3}_{n_3} (y_3) \int_{S^5} d \xi \, Y^{I_1}_{n_1} Y^{I_2}_{n_2} Y^{I_3}_{n_3}\,,
\eeq
which were computed in \cite{Arutyunov:1999en}
\beq
\int_{S^5} d \xi \, Y^{I_1}_{n_1} Y^{I_2}_{n_2} Y^{I_3}_{n_3}
= S_{n_1,n_2,n_3} \< C^{I_1}_{n_1} C^{I_2}_{n_2} C^{I_3}_{n_3} \>\,,
\label{sphere_integral}
\eeq
where $\< C^{I_1}_{n_1} C^{I_2}_{n_2} C^{I_3}_{n_3} \>$ is the unique $SO(6)$ invariant that can be formed out of three copies of $C^I_{i_1 \ldots i_n}$ and
\bea
S_{n_1,n_2,n_3} &= \frac{\omega_5}{2^{\tl\Sigma - 1}} \frac{n_1! n_2! n_3! z(n_1) z(n_2) z(n_3)}{(\tl\Sigma+2)!\tl\a_{123}! \tl\a_{231}! \tl\a_{312}!}\,,\\
\tl\Sigma&=\tfrac12(n_1+n_2+n_3) \,, \qquad \tl\alpha_{ijk} = \tfrac12 (n_i+n_j-n_k) \,.
\eea{S_def}
Furthermore, the completeness relation \eqref{completeness} implies the following relation to the four-point partial waves \eqref{eq:Z}
\begin{align}
{}&C^{I_1}_{n_1} (y_1) C^{I_2}_{n_2} (y_2) C^{I_3}_{n_3} (y_3) C^{I_4}_{n_4} (y_4)
\< C^{I_1}_{n_1} C^{I_2}_{n_2} C^{I}_{n_5} \> \< C^{I}_{n_5} C^{I_3}_{n_3} C^{I_4}_{n_4} \>\nonumber\\
={}& \frac{1}{(n_5!)^2} (y_1\cdot y_2)^{\tl\a_{125}} (y_1\cdot \partial_x)^{\tl\a_{152}} (y_2\cdot \partial_x)^{\tl\a_{251}} (y_3\cdot y_4)^{\tl\a_{345}} (y_3\cdot \partial_{\xb})^{\tl\a_{354}} (y_4\cdot \partial_{\xb})^{\tl\a_{453}}
\pi^{(6)}_{n_5}(x,\xb)\nonumber\\
={}&  (y_1\cdot y_2)^{\frac{n_1+n_2}{2}} (y_3\cdot y_4)^{\frac{n_3+n_4}{2}} \left( \frac{y_2\cdot y_4}{y_1\cdot y_4} \right)^\frac{n_2-n_1}{2} \left( \frac{y_1\cdot y_3}{y_1\cdot y_4} \right)^\frac{n_3-n_4}{2} Z_{0,n_5}^{r,s} (\al,\ab)\,,
\label{3pt_to_4pt}
\end{align}
which can be checked using \eqref{projector}.
Using this relation, the observation \eqref{h} that the expansion coefficients of the function $P(\a,\ab)$ can be written in terms of the coefficients from \eqref{sphere_integral} means
that $P(\a,\ab)$ is just the sum over products of two sphere overlaps
\bea
{}&(y_1\cdot y_2)^{\frac{p_1+p_2-4}{2}} (y_3\cdot y_4)^{\frac{p_3+p_4-4}{2}} \left( \frac{y_2\cdot y_4}{y_1\cdot y_4} \right)^\frac{p_2-p_1}{2} \left( \frac{y_1\cdot y_3}{y_1\cdot y_4} \right)^\frac{p_3-p_4}{2}
 P(\al,\ab)\\
&= \prod\limits_{i=1}^4 \left(\frac{C^{I_i}_{p_i-2} (y_i)}{ \sqrt{\Gamma(p_i)\Gamma(p_i-1)}}\right)
\frac{1}{\omega_5^2} \sum\limits_n \int_{S^5} d \xi_1 \, Y^{I_1}_{p_1-2} Y^{I_2}_{p_2-2} Y^{I}_{n}
\int_{S^5} d \xi_2 \, Y^{I_3}_{p_3-2} Y^{I_4}_{p_4-2} Y^{I}_{n}\,.
\eea{P_vs_overlaps}

\bibliographystyle{JHEP}
\bibliography{fsl}
\end{document}